\documentclass[12pt]{article}

\begin{document}
\title{NONADIABATIC CHARGED SPHERICAL GRAVITATIONAL COLLAPSE}
\author{A. Di Prisco$^1$\thanks{e-mail: adiprisc@fisica.ciens.ucv.ve}, L.
Herrera$^1$\thanks{e-mail: laherrera@cantv.net.ve}, G. Le
Denmat$^{2,3}$\thanks{%
e-mail: gele@ccr.jussieu.fr}, M. A. H. MacCallum$^4$\thanks{%
e-mail: M.A.H.MacCallum@qmul.ac.uk}\\
and N. O. Santos$^{4,5}$\thanks{%
e-mail: N.O.Santos@qmul.ac.uk} \\
{\small $^1$Escuela de F\'{\i}sica, Facultad de Ciencias,}\\
{\small Universidad Central de Venezuela, Caracas, Venezuela.}\\
{\small $^2$Universit\'e Pierre et Marie Curie - CNRS/UMR 8112,}\\
{\small LERMA/ERGA, Bo\^{\i}te 142, 4 place Jussieu, 75252 Paris
Cedex 05,
France.}\\
{\small $^3$ Observatoire de Paris,
France.}\\
{\small $^4$School of Mathematical Sciences, Queen Mary,}\\
{\small University of London, London E1 4NS, UK.}\\
{\small $^5$Laborat\'orio Nacional de Computa\c{c}\~ao Cient\'{\i}fica,}\\
{\small 25651-070 Petr\'opolis RJ, Brazil.}}
\maketitle

\newpage
\begin{abstract}
We present a complete set of the equations and  matching conditions
required for the description of physically meaningful charged,
dissipative, spherically symmetric gravitational collapse with
shear. Dissipation is described with both free-streaming and diffusion
approximations. The effects of viscosity are also
taken into account. The roles of different terms in the dynamical
equation are analyzed in detail. The dynamical equation is
coupled to a causal transport equation in  the context of
Israel-Stewart theory.  The decrease of the inertial mass
density of the fluid, by a factor which depends  on its internal
thermodynamic state, is reobtained, with the viscosity terms
included.  In accordance with the equivalence principle, the same
decrease factor is obtained for the gravitational force term.
The effect of the electric charge on the relation between the Weyl
tensor and the inhomogeneity of energy density is discussed. 
\end{abstract}

\newpage
\section{Introduction}
The study of self-gravitating spherically symmmetric charged fluid
distributions has a long and a venerable history, starting with
Rosseland and Eddington's contributions \cite{Rosseland,Eddington}.
Since then a large number of works have been dedicated to making
manifest the influence of electric charge on the structure and
evolution of self-gravitating systems (see
\cite{Rosseland}-\cite{Barreto} and references therein).

Although  some of these works refer to static situations
(\cite{Majumdar}-\cite{Bonnor},
\cite{de,Cooper,Zheng,HJPL,Treves,Felice},
\cite{Mak}-\cite{Anninos}, \cite{Ray,Cuesta,Marko,Bomer}) there
have been important efforts in describing dynamical situations too
(\cite{Israel}-\cite{Novikov}, \cite{Bardeen}-\cite{Bek},
\cite{Boulware,Banerje,Chakra,partovi,medina,Ghezzi,Barreto}).
Particularly relevant for the present paper are references
\cite{Bek,partovi} and \cite{Ghezzi}.

A renewed interest in this subject emerges from the appearance of new
mechanisms allowing for the presence of huge electric charge in
self-gravitating systems. From simple classical considerations, it can
be shown that physical objects with large amounts of charge (much
larger than 100 Coulomb per solar mass) cannot exist
\cite{Rosseland,Eddington,Glendenning}. Furthermore, as shown by
Bekenstein \cite{Bek} the electric charge is bounded by the fact that
the resulting electric field should not exceed the critical field for
pair creation, $10^{16}$ V cm$^{-1}$.
However, these restrictions have been questioned by several
authors \cite{VF,Olson,Bally,Cuesta}. Particularly appealing is
the possibility of very high electric fields in strange stars with
quark matter (see \cite{Usov,Marko} and references therein).

All this having been said, it should be clear that the restrictions
mentioned above refer to equilibrium (stable) configurations. They
do not apply to phases of intense dynamical activity with
time scales of the order of (or even smaller than)
the hydrostatic time scale, and  for which the quasi-static
approximation is clearly not reliable (e.g.\ the collapse of very
massive stars \cite{Iben} or the quick collapse phase preceding
neutron star formation, see for example \cite{myra} and references
therein). The description of this very dynamic regime is the main
purpose of this manuscript.

Besides electric charge, which will be assumed to comove with the
fluid, we shall also consider dissipative phenomena.
It is already an established fact that gravitational collapse is
a highly dissipative process (see \cite{Hs}-\cite{Mitra} and
references therein), so
the relevance of dissipation in its study cannot be over-emphasized.
Dissipation due to the emission of massless particles, photons,
and/or neutrinos is a characteristic process in the evolution of
massive stars. In fact, it seems that the only plausible mechanism
for carrying away the bulk of the binding energy of a star collapsing
to a neutron star or black hole is neutrino emission \cite{1}.

In the diffusion approximation, it is assumed that the energy flux
of radiation, like that of thermal conduction, is proportional to
the gradient of temperature. This assumption is in general very
sensible, since the mean free path of particles responsible for
the propagation of energy in stellar interiors is normally very
small compared with the typical length of the object. Thus, for
a main sequence star such as the sun, the mean free path of photons at
the centre is of the order of $2$ cm. Also, the mean free path of
trapped neutrinos in compact cores of densities above about $10^{12}$ g
cm$^{-3}$ becomes smaller than the size of the stellar core
\cite{3,4}.

Furthermore, the observational data collected from supernova 1987A
indicates that the regime of radiation transport prevailing during the
emission process is closer to the diffusion approximation than to the
free streaming limit \cite{5}.

However, in many other circumstances the mean free path of particles
transporting energy may be large enough to justify the free streaming
approximation. Therefore it is advisable to include simultaneously
both limiting cases of radiative transport, diffusion and free
streaming, allowing us, by taking both in combination, to describe
a wide range of situations.

The effects of dissipation, in both limiting cases of radiative
transport, within the context of the quasi--static approximation,
have been studied in  \cite{6'}. Using this approximation is very sensible
because the hydrostatic time scale is very small, compared with stellar
lifetimes, for many phases
of the life of a star. It is of the order of 27 minutes for the
sun, 4.5 seconds for a white dwarf and $10^{-4}$ seconds for a
neutron star of one solar mass and $10$ km radius \cite{7'}.
 However, such an approximation does not apply to the very dynamic phases
mentioned before. In those cases it is mandatory to take into account
terms which describe departure from
equilibrium, i.e.\ a full dynamic description has to be used.

For the sake of generality, we have considered a locally anisotropic
fluid. In fact, the assumption of local anisotropy of pressure,
which seems to be very reasonable for describing the matter distribution
under a variety of circumstances, has been proved to be very
useful in the study of relativistic compact objects (see
\cite{Hetal,PR}  and references therein).

Finally we have also included viscous effects in our study. In
fact, though they are generally excluded in general relativistic
models of stars, they are known to play a very important role in
the structure and evolution of neutron stars. Indeed, depending on
the dominant process, the coefficient of shear viscosity may be as
large as $\eta \approx 10^{20}$ g cm$^{-1}$ s$^{-1}$ (see
\cite{Anderson} for a review on shear viscosity in neutron stars).
Also, a theorem by Raychaudhuri and De \cite{De}, which states
that in the evolution of non-dissipative charged dust the shear
cannot vanish, emphazises the relevance of the shear in the
evolution of charged fluids.

On the other hand the coefficient of bulk viscosity may be as
large as $10^{30}$ g cm$^{-1}$ s$^{-1}$ due to Urca processes in
strange quark matter \cite{sad} (see also \cite{Dong} for a review
on bulk viscosity in nuclear and quark matter).

\section{The  energy-momentum tensor and the field equations}
In this section we  provide a full description of the matter distribution,
the line element, both inside and outside
the fluid boundary, and the field equations this line element
must satisfy.

\subsection{Interior spacetime}
We consider a spherically symmetric distribution  of collapsing
charged fluid, bounded by a spherical surface $\Sigma$: we
assume the fluid to be locally anisotropic and undergoing dissipation in the
form of heat flow, free streaming radiation and shearing
viscosity. For short we call this `matter'.
Choosing comoving coordinates inside $\Sigma$, the general
interior metric can be written
\begin{equation}
ds^2_-=-A^2dt^2+B^2dr^2+(Cr)^2(d\theta^2+\sin^2\theta d\phi^2),
\label{1}
\end{equation}
where $A$, $B$ and $C$ are functions of $t$ and $r$ and are assumed
positive. We number the coordinates $x^0=t$, $x^1=r$, $x^2=\theta$
and $x^3=\phi$.

The assumed matter energy momentum $T_{\alpha\beta}^-$ inside $\Sigma$
has the form
\begin{equation}
T_{\alpha\beta}^-=(\mu +
P_{\perp})V_{\alpha}V_{\beta}+P_{\perp}g_{\alpha\beta}+(P_r-P_{\perp})\chi_{
\alpha}\chi_{\beta}+q_{\alpha}V_{\beta}+V_{\alpha}q_{\beta}+
\epsilon l_{\alpha}l_{\beta}-2\eta\sigma_{\alpha\beta}, \label{3}
\end{equation}
where $\mu$ is the energy density, $P_r$ the radial pressure,
$P_{\perp}$ the tangential pressure, $q^{\alpha}$ the heat flux,
$\epsilon$ the radiation density, $\eta$ the coefficient of
shear viscosity, $V^{\alpha}$ the four velocity of the fluid,
$\chi^{\alpha}$ a unit four vector along the radial direction
and $l^{\alpha}$ a radial null four vector. These quantities
satisfy
\begin{equation}
V^{\alpha}V_{\alpha}=-1, \;\; V^{\alpha}q_{\alpha}=0, \;\;
\chi^{\alpha}\chi_{\alpha}=1, \;\; \chi^{\alpha} V_{\alpha}=0,
\;\; l^{\alpha} V_{\alpha}=-1, \;\; l^{\alpha}l_{\alpha}=0, \label{4}
\end{equation}
and the shear $\sigma_{\alpha\beta}$ is given by
\begin{equation}
\sigma_{\alpha\beta}=V_{(\alpha
;\beta)}+a_{(\alpha}V_{\beta)}-\frac{1}{3}\Theta(g_{\alpha\beta}+V_{\alpha}V
_{\beta}),
\label{4a}
\end{equation}
where the acceleration $a_{\alpha}$ and the expansion $\Theta$ are
given by
\begin{equation}
a_{\alpha}=V_{\alpha ;\beta}V^{\beta}, \;\;
\Theta={V^{\alpha}}_{;\alpha}. \label{4b}
\end{equation}

We do not explicitly add bulk viscosity to the system because it
can be absorbed into the radial and tangential pressures, $P_r$ and
$P_{\perp}$, of the
collapsing fluid \cite{Chan}.

It should be noted that for a physically meaningful specific model we
would need constitutive equations which would relate and determine the
quantities $\mu$, $P_r$, $P_{\perp}$, $q^{\alpha}$, $\epsilon$ and
$\eta$. Without such relations we still have so many free functions
that nothing useful can be said about the behaviour of an individual
case. However, we will show that some important general physical
results follow just from assuming, for example, that dissipation
carries energy radially outwards.

Since we assumed the metric (\ref{1}) comoving then
\begin{equation}
V^{\alpha}=A^{-1}\delta_0^{\alpha}, \;\;
q^{\alpha}=qB^{-1}\delta^{\alpha}_1, \;\;
l^{\alpha}=A^{-1}\delta^{\alpha}_0+B^{-1}\delta^{\alpha}_1, \;\;
\chi^{\alpha}=B^{-1}\delta^{\alpha}_1, \label{5}
\end{equation}
where $q$ is a function of $t$ and $r$. With (\ref{5}) we obtain
for (\ref{4a}) its non null components
\begin{equation}
\sigma_{11}=\frac{2}{3}B^2\sigma, \;\;
\sigma_{22}=\frac{\sigma_{33}}{\sin^2\theta}=-\frac{1}{3}(Cr)^2\sigma,
 \label{5a}
\end{equation}
where
\begin{equation}
\sigma=\frac{1}{A}\left(\frac{\dot{B}}{B}-\frac{\dot{C}}{C}\right),\label{5b1}
\end{equation}
and the dot stands for differentiation with respect to $t$, which
gives the scalar quantity
\begin{equation}
\sigma_{\alpha\beta}\sigma^{\alpha\beta}=\frac{2}{3}\sigma^2.
\label{5b}
\end{equation}
For
(\ref{4b}) with (\ref{5}) we have,
\begin{equation}
a_1=\frac{A^{\prime}}{A}, \;\;
\Theta=\frac{1}{A}\left(\frac{\dot{B}}{B}+2\frac{\dot{C}}{C}\right),
\label{5c}
\end{equation}
where the  prime stands for $r$
differentiation.

\subsection{The electromagnetic energy tensor and the Maxwell equations}
The electromagnetic energy tensor $E_{\alpha\beta}^-$ is given by
\begin{equation}
E_{\alpha\beta}^-=\frac{1}{4\pi}\left({F_{\alpha}}^{\gamma}F_{\beta\gamma}
-\frac{1}{4}F^{\gamma\delta}F_{\gamma\delta}g_{\alpha\beta}\right),
\label{6}
\end{equation}
where $F_{\alpha\beta}$ is the electromagnetic field tensor.
Maxwell's equations can be written
\begin{eqnarray}
F_{\alpha\beta}=\phi_{\beta,\alpha}-\phi_{\alpha,\beta}, \label{7}
\\
{F^{\alpha\beta}}_{;\beta}=4\pi J^{\alpha}, \label{8}
\end{eqnarray}
where $\phi_{\alpha}$ is the four potential and $J_{\alpha}$ is
the four current. Since the charge is assumed to be at rest with respect to the
coordinate system used in (\ref{1}), there is no magnetic field
present in this local coordinate system, and therefore we can
write
\begin{equation}
\phi_{\alpha}=\Phi\delta_{\alpha}^0, \;\; J^{\alpha}=\varsigma
V^{\alpha}, \label{9}
\end{equation}
where $\varsigma$, the charge density, and $\Phi$ are both functions
of $t$ and $r$. Charge conservation implies that
\begin{equation}
s(r)=4\pi\int^r_0\varsigma B(Cr)^2dr, \label{13}
\end{equation}
which is the electric charge interior to radius $r$, is time-independent.

With (\ref{1}) and (\ref{5}) we obtain for the
Maxwell equations (\ref{7}) and (\ref{8})
\begin{eqnarray}
\Phi^{\prime\prime}-\left(\frac{A^{\prime}}{A}+\frac{B^{\prime}}{B}-2\frac{C
^{\prime}}{C}-\frac{2} {r}\right)\Phi^{\prime}=4\pi\varsigma
AB^2, \label{10} \\
\dot \Phi^{\prime}-\left(\frac{\dot{A}
}{A}+\frac{\dot{B}}{B}-2\frac{\dot{C}}{C}\right)\Phi^{\prime}=0.
\label{11}
\end{eqnarray}
Integrating (\ref{10}) and (\ref{11}) produces
\begin{equation}
\Phi^{\prime}=\frac{sAB}{(Cr)^2}. \label{12}
\end{equation}

\subsection{The Einstein equations}
Einstein's field equations for the interior spacetime (\ref{1}) are given by
\begin{equation}
G_{\alpha\beta}^-=8\pi(T_{\alpha\beta}^-+E_{\alpha\beta}^-).
\label{2}
\end{equation}

The non null components of (\ref{2})
with (\ref{1}), (\ref{3}), (\ref{5}), (\ref{6}) and (\ref{12})
become
\begin{eqnarray}
8\pi(T_{00}^-+E_{00}^-)&=&8\pi(\mu+\epsilon)A^2+\frac{(sA)^2}{(Cr)^4}
\nonumber \\
&=&\left(2\frac{\dot{B}}{B}+\frac{\dot{C}}{C}\right)\frac{\dot{C}}{C}
 +\left(\frac{A}{B}\right)^2\left\{-2\frac{C^{\prime\prime}}{C}
 +\left(2\frac{B^{\prime}}{B}
 -\frac{C^{\prime}}{C}\right)\frac{C^{\prime}}{C} \right.\nonumber\\
&&\left.+\frac{2}{r}\left(\frac{B^{\prime}}{B}-3\frac{C^{\prime}}{C}\right)
  -\left[1-\left(\frac{B}{C}\right)^2\right]\frac{1}{r^2}\right\},
  \label{14} \\
8\pi(T_{01}^-+E_{01}^-)&=&-8\pi(q+\epsilon)AB\nonumber\\
&=&-2\left(\frac{\dot{C}^{\prime}}{C}-\frac{\dot{B}}{B}\frac{C^{\prime}}{C}
 -\frac{\dot{C}}{C}\frac{A^{\prime}}{A}\right)
 +\frac{2}{r}\left(\frac{\dot{B}}{B}-\frac{\dot{C}}{C}\right),
 \label{15} \\
8\pi(T_{11}^-+E_{11}^-)&=&
 8\pi\left(P_r+\epsilon-\frac{4}{3}\eta\sigma\right)B^2
 -\frac{(sB)^2}{(Cr)^4}\nonumber \\
&=&-\left(\frac{B}{A}\right)^2\left[2\frac{\ddot{C}}{C}
 +\left(\frac{\dot{C}}{C}\right)^2
 -2\frac{\dot{A}}{A}\frac{\dot{C}}{C}\right] \label{16b}\\
&&+\left(\frac{C^{\prime}}{C}\right)^2
 +2\frac{A^{\prime}}{A}\frac{C^{\prime}}{C}
 +\frac{2}{r}\left(\frac{A^{\prime}}{A}
 +\frac{C^{\prime}}{C}\right)
 +\left[1-\left(\frac{B}{C}\right)^2\right]\frac{1}{r^2},
 \nonumber\\
8\pi(T_{22}^-+E_{22}^-)&=&\frac{8\pi}{\sin^2\theta}(T_{33}^-+E_{33}^-)
 \nonumber \\
&=&8\pi\left(P_{\perp}+\frac{2}{3}\eta\sigma\right)(Cr)^2
 +\left(\frac{s}{Cr}\right)^2 \nonumber \\
&=&-\left(\frac{Cr}{A}\right)^2\left[\frac{\ddot{B}}{B}+\frac{\ddot{C}}{C}
 -\frac{\dot{A}}{A}\left(\frac{\dot{B}}{B}+\frac{\dot{C}}{C}\right)
 +\frac{\dot{B}}{B}\frac{\dot{C}}{C}\right]\nonumber\\
&&+\left(\frac{Cr}{B}\right)^2\left[\frac{A^{\prime\prime}}{A}
 +\frac{C^{\prime\prime}}{C}-\frac{A^{\prime}}{A}\left(\frac{B^{\prime}}{B}
 -\frac{C^{\prime}}{C}\right)-\frac{B^{\prime}}{B}\frac{C^{\prime}}{C}\right.
 \nonumber \\
&& \left.+\frac{1}{r}\left(\frac{A^{\prime}}{A}-\frac{B^{\prime}}{B}
 +2\frac{C^{\prime}}{C}\right)\right].\label{17}
\end{eqnarray}

The component (\ref{15}) can be rewritten with (\ref{5b1}) and
(\ref{5c}) as
\begin{equation}
4\pi(q+\epsilon)B=\frac{1}{3}(\Theta-\sigma)^{\prime}
-\sigma\frac{(Cr)^{\prime}}{Cr}.\label{17a}
\end{equation}

Next, the mass function $m(t,r)$ introduced by Misner and Sharp
\cite{Misner} (see also \cite{Cahill}) can be generalized to include the
electromagnetic contribution by

\begin{equation}
m=\frac{(Cr)^3}{2}{R_{23}}^{23}+\frac{s^2}{2Cr}
=\frac{Cr}{2}\left\{\left(\frac{r\dot{C}}{A}\right)^2
-\left[\frac{(Cr)^{\prime}}{B}\right]^2+1\right\}+\frac{s^2}{2Cr},
 \label{18}
\end{equation}
which is the same mass function used in \cite{Bek} and \cite{partovi}.

\subsection{The exterior spacetime and junction conditions}
Outside $\Sigma$ we assume we have the Reissner-Nordstr\"om-Vaidya
spacetime (i.e.\ we assume all outgoing radiation is massless),
described by
\begin{equation}
ds^2=-\left[1-\frac{2M(v)}{r}+\frac{Q^2}{r^2}\right]dv^2-2drdv+r^2(d\theta^2
+r^2\sin\theta^2
d\phi^2) \label{1int}
\end{equation}
where $M(v)$ and $Q$ denote the total mass and charge respectively,
and  $v$ is the retarded time.

The junction conditions for the smooth  matching of an adiabatic
charged sphere to the Reissner-Nordstr\"om spacetime were
discussed in detail in \cite{partovi} and \cite{Ghezzi}, whereas the
matching of the full non-adiabatic sphere  (including viscosity) to
the Vaidya spacetime was discussed in
\cite{chan1}. The result is that the matching of
(\ref{1})  and (\ref{1int}) on $\Sigma$ implies
\begin{equation}
P_r+4\eta \sigma\stackrel{\Sigma}{=}q, \quad
m(t,r)\stackrel{\Sigma}{=}M(v), \quad s \stackrel{\Sigma}{=} Q\label{junction},
\end{equation}
where $\stackrel{\Sigma}{=}$ means that both sides of the equation
are evaluated on $\Sigma$.

\section{Dynamical equations}

The non trivial components of the Bianchi identities ,
$(T^{-\alpha\beta}+E^{-\alpha\beta})_{;\beta}=0$, from (\ref{3}) and
(\ref{6}) yield
\begin{eqnarray}
(T^{-\alpha\beta}+E^{-\alpha\beta})_{;\beta}V_{\alpha}
 &=&-\frac{1}{A}(\dot{\mu}+\dot{\epsilon})
   -\left(\mu+P_r+2\epsilon-\frac{4}{3}\eta\sigma\right)
    \frac{\dot B}{AB}\nonumber\\
&&-2\left(\mu+P_{\perp}+\epsilon+\frac{2}{3}\eta\sigma\right)
  \frac{\dot C}{AC}\label{II18}\\
&&-\frac{1}{B}(q+\epsilon)^{\prime}-2(q+\epsilon)\frac{(ACr)^{\prime}}{ABCr}
 \phantom{-\frac{1}{A}(\dot{\mu}+\dot{\epsilon})}=0,\nonumber \\
(T^{-\alpha\beta}+E^{-\alpha\beta})_{;\beta}\chi_{\alpha}
&=&\frac{1}{A}(\dot{q}+\dot{\epsilon})
+\frac{1}{B}\left(P_r+\epsilon-\frac{4}{3}\eta\sigma\right)^{\prime}
 \nonumber\\
&&+2(q+\epsilon)\frac{\dot B}{AB}+2(q +\epsilon)\frac{\dot C}{AC} \nonumber\\
&&
 +\left(\mu+P_r+2\epsilon-\frac{4}{3}\eta\sigma\right)\frac{A^{\prime}}{AB}
\label{III18}\\
&&+2(P_r-P_{\perp}+\epsilon-2\eta\sigma)\frac{(Cr)^{\prime}}{BCr}
 -\frac{ss^{\prime}}{4\pi B(Cr)^4}=0,  \nonumber
\end{eqnarray}
where we used (\ref{12}) and (\ref{13}).

To study the dynamical properties of the system, let us  introduce,
following Misner and Sharp \cite{Misner}, the proper time derivative $D_T$
given by
\begin{equation}
D_T=\frac{1}{A}\frac{\partial}{\partial t}, \label{16}
\end{equation}
and the proper radial derivative $D_R$,
\begin{equation}
D_R=\frac{1}{R^{\prime}}\frac{\partial}{\partial r}, \label{23a}
\end{equation}
where
\begin{equation}
R=Cr, \label{23aa}
\end{equation}
defines the proper radius of a spherical surface inside $\Sigma$, as
measured from its area.

Using (\ref{16}) we can define the velocity $U$ of the collapsing
fluid as the variation of the proper radius with respect to proper time, i.e.\
\begin{equation}
U=rD_TC<0 \;\; \mbox{(in the case of collapse)}. \label{19}
\end{equation}
Then (\ref{18}) can be rewritten as
\begin{equation}
E \equiv \frac{(Cr)^{\prime}}{B}=\left[1+U^2-\frac{2m(t,r)}{Cr}
+\left(\frac{s}{Cr}\right)^2\right]^{1/2}.
\label{20}
\end{equation}
With (\ref{23a})-(\ref{23aa}) we can express (\ref{17a}) as
\begin{equation}
4\pi(q+\epsilon)=E\left[\frac{1}{3}D_R(\Theta-\sigma)
-\frac{\sigma}{R}\right].\label{21a}
\end{equation}

Using (\ref{14}-\ref{17}) and (\ref{16}-\ref{23aa}) we obtain from
(\ref{18})
\begin{eqnarray}
D_Tm=-4\pi\left[\left
(P_r+\epsilon-\frac{4}{3}\eta\sigma\right)U+(q+\epsilon)E\right]R^2,
\label{22}
\end{eqnarray}
and
\begin{eqnarray}
D_Rm=4\pi\left[\mu+\epsilon+(q+\epsilon)\frac{U}{E}\right]R^2+\frac{s}{R}D_Rs.
\label{27}
\end{eqnarray}
Expression  (\ref{22}) describes the rate of variation of the
total energy inside a surface of radius $Cr$. On the right hand
side of (\ref{22}), $(P_r+\epsilon-4\eta\sigma/3)U$ (in the case
of collapse $U<0$) increases the energy inside $Cr$ through the
rate of work being done by the ``effective'' radial pressure
$P_r-4\eta\sigma/3$ and the radiation pressure $\epsilon$. Clearly
here the heat flux $q$ does not appear since there is no
pressure associated with the diffusion process. The second term
$-(q+\epsilon)E$ is the matter energy leaving the spherical
surface.

Equation (\ref{27}) shows how the total energy enclosed varies between
neighboring spherical surfaces inside the
fluid distribution.
The first term on the right hand side of (\ref{27}), $\mu+\epsilon$, is due
to the energy density of the fluid element plus the energy density of the
null fluid describing dissipation
in the free streaming approximation. The second term,
$(q+\epsilon)U/E$ is negative (in the case of collapse) and measures the
outflow of heat and radiation. Finally the last term is the electrostatic
contribution.

Equation (\ref{27}) may be integrated to obtain
\begin{equation}
m=\int^{R}_{0}4\pi R^2 \left[\mu +
\epsilon+(q+\epsilon)\frac{U}{E}\right]dR+\frac{s^2}{2R}+\frac{1}{2}\int^{R
}_{0}\frac{s^2}{R^2}dR
\label{27int}
\end{equation}
(assuming a regular centre to the distribution, so $m(0)=0$).

The acceleration $D_TU$ of an infalling particle inside $\Sigma$ can
be obtained by using (\ref{16b}), (\ref{16}), (\ref{18}) and (\ref{20}),
producing
\begin{equation}
D_TU=-\frac{m}{R^2}-4\pi\left(P_r+\epsilon-\frac{4}{3}\eta\sigma\right)R
+\frac{s^2}{R^3}+\frac{EA^{\prime}}{AB}, \label{28}
\end{equation}
and then, substituting $A^{\prime}/A$ from (\ref{28}) into
(\ref{III18}), we obtain 
\begin{equation}
\arraycolsep=0pt
\begin{array}{rl}
\left(\mu+P_r+2\epsilon-\frac{4}{3}\eta\sigma\right) D_TU \phantom{lll}
 &\\
=\phantom{l}-\left(\mu+P_r+2\epsilon-\frac{4}{3}\eta\sigma\right)&
\displaystyle{\left[\frac{m}{R^2}
+4\pi\left(P_r+\epsilon-\frac{4}{3}\eta\sigma\right)R
-\frac{s^2}{R^3}\right]}\\ 
-E^2\left[D_R\left(P_r+\epsilon-\frac{4}{3}\eta\sigma\right)\right.&
+\displaystyle{2\left.(P_r-P_{\perp}+\epsilon-2\eta\sigma)\frac{1}{R}
-\frac{s}{4\pi R^4}D_Rs\right]}\\ 
-E&\left[D_Tq+D_T\epsilon+4(q+\epsilon)\frac{U}{R}
 +2(q+\epsilon)\sigma\right], \end{array}
\label{3m}
\end{equation}
which in the non-dissipative locally isotropic case coincides
with Eq.\ (43) in \cite{Bek}. Let us now analyze in some detail
the three terms on the right of (\ref{3m}).

The first term  on the right hand side of (\ref{3m}) represents the
gravitational force. The factor within the round bracket (the same factor
as on the left of (\ref{3m})) defines the
inertial mass density (``passive''  gravitational mass density) and shows
how it is affected by dissipative terms. Observe that it is not affected by
the electric charge.

The factor within
the first square bracket shows how dissipation and the electric charge
affect the ``active'' gravitational mass term.  Using (\ref{27int}) in
(\ref{3m}) we see that the charge
will increase the ``active gravitational mass'' only if
\begin{equation}
\int^{R}_{0}\frac{s^2}{R^2}dR>\frac{s^2}{R}
\label{con}
\end{equation}
or, equivalently
\begin{equation}
\frac{s}{R}>D_Rs;
\label{con1}
\end{equation}
otherwise it will decrease it. This strange effect was already noticed by
Bekenstein \cite{Bek}, and enhances the possibility that  Coulomb repulsion
might prevent the gravitational
collapse of the sphere.

There are three different contributions in the second square bracket. The
first one is just the gradient of the total ``effective'' radial  pressure
(which includes the radiation
pressure and the influence of shear viscosity on $P_r$). The second
contribution comes from the local anisotropy of pressure, including the
contributions from the radiation pressure
and shear viscosity. Finally the last term describes Coulomb repulsion,
which is always positive (always opposing gravitation).

The last square bracket contains different contributions due to dissipative
processes. The third term within this bracket is positive ($U<0$) showing
that the outflow of
$q>0$ and $\epsilon>0$ diminish the total energy inside the collapsing
sphere, thereby reducing the rate of collapse. The last term describes an
effect resulting from the coupling of
the dissipative flux with the shear of the fluid. The effects of
$D_T\epsilon$ have
been discussed in detail in
\cite{MisnerII}. Thus it only remains to analyse the effects of $D_Tq$;
this depends on the transport equation adopted, and we will proceed to
study one case in the next section.

However before doing that it is instructive to recover  a known result for
the static case.

\subsection{ Static charged dust}
In the limit of hydrostatic equilibrium when $U=\sigma=q=\epsilon=0$, we
have from (\ref{III18})
\begin{equation}
P^{\prime}_r+(\mu+P_r)\frac{A^{\prime}}{A}+2(P_r-P_{\perp})\frac{(Cr)^{\prime}}{Cr}-\frac{ss^{\prime}}{4\pi(Cr)^4}=0,
\label{hh1}
\end{equation}
which is just the generalization of the Tolman-Oppenheimer-Volkov
equation for anisotropic charged fluids obtained in \cite{Chan1}
while studying dynamical instability for radiating anisotropic
collapse.

When the static fluid reduces to charged dust, with
$P_r=P_{\perp}=0$, then,  by using (\ref{13}), (\ref{hh1}) becomes
\begin{equation}
\mu\frac{A^{\prime}}{A}-\frac{s\varsigma B}{(Cr)^2}=0. \label{hh2}
\end{equation}
Since $B$ and $C$ depend only on $r$, we can transform $r$ so that
 $B=C$. Eliminating $s$ from  the field equations (\ref{16b}) and
 (\ref{17}),
we can solve for $AB$, and imposing regularity conditions and rescaling
 $t$ we have
\begin{equation}
C=B, \;\; AB=1, \;\; s^2=r^4B^{\prime 2}. \label{hh3}
\end{equation}
Substituting (\ref{hh3}) into (\ref{hh2}) we obtain
\begin{equation}
\mu^2=\varsigma^2, \label{hh4}
\end{equation}
which is the well known result originally obtained by Bonnor
\cite{Bonnor} for arbitrary symmetry.

\section{The transport equation}
 We shall use a transport equation derived from the
M\"{u}ller-Israel-Stewart second
order phenomenological theory for dissipative fluids \cite{Muller67, IsSt76}.

Indeed, it is well known that the Maxwell-Fourier law for
radiation flux leads to a parabolic equation (diffusion equation)
which predicts propagation of perturbations with infinite speed
(see \cite{6}-\cite{8'} and references therein). This simple fact
is at the origin of the pathologies \cite{9} found in the
approaches of Eckart \cite{10} and Landau \cite{11} for
relativistic dissipative processes. To overcome such difficulties,
various relativistic
theories with non-vanishing relaxation times have been proposed in
the past \cite{Muller67,IsSt76,14,15}. The important point is that
all these theories provide a heat transport equation which is not
of Maxwell-Fourier type but of Cattaneo type \cite{18}, leading
thereby to a hyperbolic equation for the propagation of thermal
perturbations.

The corresponding  transport equation for the heat flux reads
\begin{equation}
\tau
h^{\alpha\beta}V^{\gamma}q_{\beta;\gamma}+q^{\alpha}=-\kappa h^{\alpha\beta}
(T_{,\beta}+Ta_{\beta}) -\frac 12\kappa T^2\left( \frac{\tau
V^\beta }{\kappa T^2}\right) _{;\beta }q^\alpha ,  \label{21t}
\end{equation}
where $h^{\mu \nu }$ is the projector onto the three space orthogonal to $%
V^\mu $, $\kappa $  denotes the thermal conductivity, and  $T$ and
$\tau$ denote temperature and relaxation time respectively. Observe
that, due to the symmetry of the problem, equation (\ref{21t}) only
has one independent component, which may be written after using
(\ref{1}), (\ref{5}) and (\ref{5c}) as
\begin{equation}
\tau{\dot q}=-\frac{1}{2}\kappa qT^2\left(\frac{\tau}{\kappa
T^2}\right)^{\dot{}}-\tau q\left(\frac{\dot
B}{2B}+\frac{\dot C}{C}\right)-\frac{\kappa}{B}(TA)^{\prime}-qA.
\label{te}
\end{equation}
Now using (\ref{16}-\ref{20}) we can rewrite (\ref{te}) as
\begin{eqnarray}
D_Tq&=&-\frac{\kappa T^2q}{2\tau}D_T\left(\frac{\tau}{\kappa T^2}\right)
     -q\left(\frac{3}{2}\frac{U}{R}
     +\frac{1}{2}\sigma+\frac{1}{\tau}\right)-\frac{\kappa E}{\tau}D_RT 
     -\frac{\kappa T}{\tau E}D_TU \nonumber\\
&&-\frac{\kappa T}{\tau E}
  \left[m+4\pi\left(P_r+\epsilon-\frac{4}{3}\eta \sigma\right)R^3
 -\frac{s^2}{R}\right]\frac{1}{R^2}.
\label{V3t}
\end{eqnarray}

We can couple the transport equation in the form above, (\ref{V3t}), to the
dynamical equation (\ref{3m}), in order to bring out
the effects of dissipation  on the dynamics of the collapsing sphere. For
that purpose,  let
us substitute (\ref{V3t}) into (\ref{3m}):  then we obtain, after some
rearrangements,
\begin{eqnarray}
\left(\mu+P_r+2\epsilon-\frac{4}{3}\sigma\eta\right)(1-\alpha)D_TU
\phantom{==}&&
 \nonumber \\
=(1-\alpha)F_{grav}+F_{hyd}+\frac{\kappa E^2}{\tau}D_RT 
&+&E\left[\frac{\kappa T^2q}{2\tau}D_T\left(\frac{\tau}{\kappa
   T^2}\right)-D_T\epsilon\right]
\nonumber \\
-Eq\left(\frac{5}{2}\frac{U}{R}+\frac{3}{2}\sigma-\frac{1}{\tau}\right)
 &-&2E\epsilon\left(2\frac{U}{R}+\sigma\right),
\label{V4}
\end{eqnarray}
where $F_{grav}$ and $F_{hyd}$ are defined by
\newpage
\begin{eqnarray}
F_{grav}&=&-\left(\mu+P_r+2\epsilon -\frac{4}{3}\eta
\sigma\right)\nonumber\\
&&\times
\left[m+4\pi\left(P_r+\epsilon-\frac{4}{3}\eta \sigma
\right)R^3-\frac{s^2}{R}\right]\frac{1}{R^2},
\label{grav}\\
F_{hyd}&=& -E^2 \left[D_R \left(P_r+\epsilon -\frac{4}{3}\eta
\sigma\right)\right. \nonumber\\
&&\left.+2(P_r-P_{\perp}+\epsilon-2\eta\sigma)\frac{1}{R}-\frac{s}{4\pi
R^4}D_Rs\right], \label{hyd}
\end{eqnarray}
and $\alpha$ is given by
\begin{equation}
\alpha=\frac{\kappa
T}{\tau}\left(\mu+P_r+2\epsilon-\frac{4}{3}\sigma\eta\right)^{-1}.
\label{alpha}
\end{equation}
Some  comments are in order at this point:
\begin{itemize}
\item Once the transport equation has been taken into account, then the
inertial energy density  and the ``passive gravitational mass density'',
i.e the factor multiplying $D_TU$ and
the first factor at the right of (\ref{3m}) respectively (which of course
are the same, as expected from the equivalence principle), appear
diminished by the factor $1-\alpha$, a result
already obtained in \cite{Hs}, but here generalized by the inclusion of
the viscosity and radiative phenomena.

\item Observe that the charge does not enter into the definition of $\alpha$.
However it does affect the ``active gravitational mass'' (the factor within
the square bracket in
(\ref{grav})).

\item The repulsive Coulomb term (the last term in (\ref{hyd})) depends on
$D_{R}s$ and always opposes gravitation. Its effect is reinforced if
$D_{R}s$ is large enough to
violate (\ref{con1}), in which  case the charge will decrease the ``active
gravitational mass'' term in (\ref{grav}).
\end{itemize}

\section{The Weyl tensor}
In this section we shall find some interesting relationships linking
the Weyl tensor with matter variables, from which we shall extract
some conclusions about the arrow of time.

{}From the Weyl tensor we may construct the Weyl scalar ${\mathcal
C}^2=C^{\alpha\beta\gamma\delta}C_{\alpha\beta\gamma\delta}$ which
can be given in terms of the Kretchman scalar ${\mathcal
R}=R^{\alpha\beta\gamma\delta}R_{\alpha\beta\gamma\delta}$, the
Ricci tensor $R_{\alpha\beta}$ and the curvature scalar R by
\begin{equation}
{\mathcal C}^2={\mathcal
R}-2R^{\alpha\beta}R_{\alpha\beta}+\frac{1}{3}\rm{R}^2.\label{I18}
\end{equation}
Substituting (\ref{a13N}) from the Appendix with the field equations
(\ref{14}-\ref{17}) into (\ref{I18}) we obtain
\begin{equation}
{\mathcal E}=m-
\frac{4\pi}{3}(\mu-P_r+P_{\perp}+2\eta\sigma)R^3-\frac{s^2}{R},\label{I19}
\end{equation}
where ${\mathcal E}$ is given by
\begin{equation}
{\mathcal E}=\frac{{\mathcal C}}{48^{1/2}}R^3, \label{I20}
\end{equation}.

{}From (\ref{I19}) with (\ref{22}) and (\ref{27}) we have
\begin{eqnarray}
D_T{\mathcal
E}&=&-4\pi\left[\frac{1}{3}R^3D_T(\mu-P_r+P_{\perp}+2\eta\sigma)\right.
 \nonumber\\
&&\phantom{-4\pi}\left.
 +\left(\mu+P_{\perp}+\epsilon+\frac{2}{3}\eta\sigma\right)
R^2U
+(q+\epsilon)ER^2\right]+\frac{s^2U}{R^2},\label{I21}
\end{eqnarray}
and
\begin{eqnarray}
D_R{\mathcal E} =4\pi\left[(q
+\epsilon)\frac{R^2 U}{E}
-\frac{1}{3}R^3D_R(\mu-P_r+P_{\perp}+2\eta\sigma)\right.\nonumber\\
\left.+(\epsilon+P_r-P_{\perp}-2\eta\sigma)R^2 \right]-\frac{sD_Rs}{R}
+\left(\frac{s}{R}\right)^2.\label{II21}
\end{eqnarray}

{}From (\ref{II21}) we obtain at once for the non--charged,
non-dissipative, perfect fluid case
\begin{equation}
D_R{\mathcal E}+\frac{4\pi}{3}R^3D_R\mu=0, \label{arrow}
\end{equation}
implying that $D_R \mu=0$ produces ${\mathcal
C}=0$ (using the regular axis condition), and conversely the
conformally flat condition implies homogeneity in
the energy density.

 This particularly  simple relation between the Weyl tensor and
density inhomogeneity, for perfect fluids, is at the origin of
Penrose's  proposal to provide a
gravitational arrow of time in terms of the Weyl tensor \cite{Pe}. The
rationale
behind this idea is that tidal forces tend to make the gravitating
fluid more inhomogeneous as the evolution proceeds, thereby
indicating the sense of time.

 However the fact that such a relationship is no longer valid in the presence
of local anisotropy of the pressure and/or
dissipative processes, already discussed in \cite{Hetal}, explains its
failure in scenarios where the above-mentioned  factors are present.
Here we see how the electric charge distribution affects the link between
the Weyl tensor and density inhomogeneity, suggesting that electric charge
(whenever present) should enter
into any definition of a gravitational arrow of time.

\section{Conclusions}
We have provided a full set of the equations required for a description of
physically meaningful models of collapsing charged spheres. We have
included dissipative phenomena as well as
anisotropic pressure; the justification for doing so was given in the
Introduction.

The role of charge distribution in the dynamics of such configurations
is clearly exhibited in equations (\ref{22}), (\ref{27}), (\ref{3m})
and (\ref{V4}). In particular it is worth stressing the fact that
electric charge, unlike pressure, does not always produce a
``regeneration effect'' (does not always increase the ``active
gravitational mass''). This fact together with the presence of the
Coulomb term in (\ref{3m}) (or (\ref{V4})) indicates the relevance of
the electric charge in the process of collapse.

Finally we have obtained a relation (\ref{II21}) exhibiting the way in
which electric charge affects the link between the Weyl tensor and density
inhomogeneity. The consequences of
this for a definition of a gravitational arrow of time have been discussed.

\section*{Acknowledgments.}
LH and ADP acknowledge financial support from the CDCH at Universidad
Central de Venezuela under grant PI 03.11.4180.1998. NOS was supported as a Visiting Fellow by EPSRC grant EP/E063896/1.
\section*{Appendix}
The spacetime (\ref{1}) has the following non null Riemann tensor
components
\begin{eqnarray}
R_{0101}&=&-B\ddot{B}+\frac{B}{A}\dot{A}\dot{B}+AA^{\prime\prime}
 -\frac{A}{B}A^{\prime}B^{\prime},\label{a1}\\
R_{0202}&=&(Cr)^2\left[-\frac{\ddot{C}}{C}+\frac{\dot{A}}{A}\frac{\dot{C}}{C}+
 \left(\frac{A}{B}\right)^2\frac{A^{\prime}}{A}
 \left(\frac{C^{\prime}}{C}+\frac{1}{r}\right)\right],\label{a2}\\
R_{0212}&=&(Cr)^2\left[-\frac{{\dot C}^{\prime}}{C}
 +\frac{\dot B}{B}\frac{C^{\prime}}{C}+\frac{{\dot C}}{C}\frac{A^{\prime}}{A}
 +\frac{1}{r}\left(\frac{\dot B}{B}-\frac{\dot C}{C}\right)\right],\label{a3}\\
R_{1212}&=&(Cr)^2\left[\left(\frac{B}{A}\right)^2\frac{\dot B}{B}
  \frac{\dot C}{C}-\frac{C^{\prime\prime}}{C}
  +\frac{B^{\prime}}{B}\frac{C^{\prime}}{C}
 +\frac{1}{r}\left(\frac{B^{\prime}}{B}-
 2\frac{C^{\prime}}{C}\right)\right],\label{a4}\\
R_{2323}&=&(Cr)^2\sin^2\theta\left[\left(\frac{r\dot C}{C}\right)^2
 -\left(\frac{rC^{\prime}}{B}\right)^2-2\frac{CrC^{\prime}}{B^2}
 -\left(\frac{C}{B}\right)^2+1\right],
\label{a5}
\end{eqnarray}
and
\begin{equation}
R_{0303}=R_{0202}\sin^2\theta, \;\; R_{0313}=R_{0212}\sin^2\theta,
\;\; R_{1313}=R_{1212}\sin^2\theta,\label{a6}
\end{equation}
hence it has 5 independent components and the Kretchman scalar
becomes
\begin{eqnarray}
{\mathcal
R}=4\left[\frac{1}{(AB)^4}(R_{0101})^2+\frac{2}{(ACr)^4}(R_{0202})^2-\frac{4
}{(AB)^2(Cr)^4}(R_{0212})^2
\right.\nonumber\\
\left.+\frac{2}{(BCr)^4}(R_{1212})^2+\frac{1}{(Cr)^8\sin^4\theta}(R_{2323})^
2\right].
\label{a7}
\end{eqnarray}
The components (\ref{a1}-\ref{a5}) can be written in terms of the
Einstein tensor $G_{\alpha\beta}=R_{\alpha\beta}-g_{\alpha\beta}\rm{R}/2$
and the mass function (\ref{18}) producing
\begin{eqnarray}
\hspace*{-1em}
R_{0101}=&(AB)^2\left[\frac{1}{2A^2}G_{00}-\frac{1}{2B^2}G_{11}
 +\frac{1}{(Cr)^2}G_{22}
 -\frac{2}{(Cr)^3}\left(m-\frac{s^2}{2Cr}\right)\right],\label{a8}\\
\hspace*{-2em}
R_{0202}=&(ACr)^2\left[\frac{1}{2B^2}G_{11}
 +\frac{1}{(Cr)^3}\left(m-\frac{s^2}{2Cr}\right)\right],\label{a9}\\
\hspace*{-2em}
R_{0212}=&\frac{(Cr)^2}{2}G_{01},\label{a10}\\
\hspace*{-2em}
R_{1212}=&(BCr)^2\left[\frac{1}{2A^2}G_{00}
 -\frac{1}{(Cr)^3}\left(m-\frac{s^2}{2Cr}\right)\right],\label{a11}\\
\hspace*{-2em}
R_{2323}=&2Cr\sin^2\theta\left(m-\frac{s^2}{2Cr}\right).\label{a12}
\end{eqnarray}
Substituting (\ref{a8}-\ref{a12}) into (\ref{a7}) we obtain
\begin{eqnarray}
{\mathcal R}&=&\frac{48}{(Cr)^6}\left(m-\frac{s^2}{2Cr}\right)^2 
 -\frac{16}{(Cr)^3}\left(m-\frac{s^2}{2Cr}\right)\left[\frac{G_{00}}{A^2}
 -\frac{G_{11}}{B^2}+\frac{G_{22}}{(Cr)^2}\right]
 \nonumber\\
&&-4\left(\frac{G_{01}}{AB}\right)^2+3\left[\left(\frac{G_{00}}{A^2}\right)^2+
 \left(\frac{G_{11}}{B^2}\right)^2\right]+
 4\left[\frac{G_{22}}{(Cr)^2}\right]^2\nonumber\\
&&-2\frac{G_{00}}{A^2}\frac{G_{11}}{B^2}
 +4\left(\frac{G_{00}}{A^2}-\frac{G_{11}}{B^2}\right)\frac{G_{22}}{(Cr)^2}.
\label{a13N}
\end{eqnarray}

\end{document}